%% file: cudatemp.tex
\documentclass[5p,authoryear,pdftex]{elsarticle}

\usepackage{amsmath,amssymb,afterpage} \usepackage{ifthen} \usepackage{comment,verbatim,url} 

\newboolean{includeversions}

\setboolean{includeversions}{false}

\newboolean{blackandwhite} \setboolean{blackandwhite}{true}

\newboolean{usepdf} \newif\ifpdf \ifx\pdfoutput\undefined \pdffalse \else \pdfoutput=1 \pdftrue \fi

\ifpdf \setboolean{usepdf}{true} \else \setboolean{usepdf}{false} \fi

\ifthenelse {\boolean{usepdf}}{ 
}{ 
} \graphicspath {{figures/}} \include {natlatex_macros} \usepackage {astrojournals}

\renewcommand{\mcrx}{{\sc sunrise}}

\journal{New Astronomy} 
\begin {document} \begin {frontmatter}

\author{Patrik Jonsson}
\ead{pjonsson@cfa.harvard.edu} \author{Joel R. Primack} \ead{joel@scipp.ucsc.edu} \address{Santa Cruz Institute for Particle Physics, University of California, Santa Cruz, CA 95064, USA} \cortext[cor1]{Corresponding author. Present address: Harvard-Smithsonian Center for Astrophysics, 60 Garden St., MS-51, Cambridge, MA 02138, USA. Tel.: +1 (617) 384-7831, fax: +1 (617) 495-7093} \ifthenelse {\boolean{includeversions}}{\address{Draft Version: \verb$Id: cudatemp.txt 1114 2009-12-23 15:37:39Z patrik $}}{}

\title {Accelerating Dust Temperature Calculations with Graphics Processing Units}


\begin{abstract}
When calculating the infrared spectral energy distributions (SEDs) of
galaxies in radiation-transfer models, the calculation of dust grain
temperatures is generally the most time-consuming part of the
calculation. Because of its highly parallel nature, this calculation is
perfectly suited for massively parallel general-purpose Graphics
Processing Units (GPUs). This paper presents an implementation of the
calculation of dust grain equilibrium temperatures on GPUs in the
Monte-Carlo radiation transfer code \mcrx , using the CUDA API. The GPU
can perform this calculation 69 times faster than the 8 CPU cores,
showing great potential for accelerating calculations of galaxy SEDs.
\end{abstract}


\end{frontmatter}


\section{Introduction}

Over the last decade, radiation-transfer models of dust in galaxies
have become increasingly capable, with higher spatial resolutions,
higher wavelength resolutions, and more realistic descriptions of the
dust itself \citep{gordonetal01, jonsson06sunrise, lietal07,
bianchi08trading, chakrabartietal07radishe, jonssonetal09}. Current
codes calculate the temperatures of grains of a number of different
compositions and sizes based on physical dust models
\citep[e.g.][]{weingartnerdraine01, draineli07} at many different
spatial locations in the problem, sometimes also including thermal
fluctuations.  As the complexity of the dust models have gone up,
researchers have found that the most computationally demanding part of
the calculation no longer is the actual transfer of radiation through
the medium, but rather the calculation of the dust emission spectrum.

At the same time, graphics-processing units (GPUs) have evolved from
specialty hardware to massively parallel general computation devices.
These devices are made to render independent pixels, where the same
computation is being done millions of times on largely independent
data. The predictable data access pattern means that the considerable
amount of silicon real estate which is used to hide memory latency on a
general-purpose CPU is unnecessary. Instead, the entire chip area can
be used for computational units, giving a raw floating-point processing
power many times larger than that of CPUs of similar cost and power
consumption.

The utility of GPUs for general high-performance computations was
initially limited by the fact that programs needed to be expressed in
terms of graphics operations and written in shading languages such as
Cg or GLSL, making it difficult to port programs.  Still, astrophysical
applications were explored, for example N-body calculations
\citep{portegieszwartetal07} with large increases in performance
compared to custom GRAPE hardware.  The programming obstacle was
largely overcome with the Nvidia CUDA \citep[Compute Unified Device
Architecture,][]{cudaguide2.3} API, which enables programs to be
written in a language very similar to C/C++, with added directives.
CUDA has been used in astrophysics not only for N-body calculations
\citep{bellemanetal08, gaburovetal09} but also for radio interferometer
data processing \citep{ordetal09} and exoplanet searches
\citep{ford09}, to name a few.

This paper presents an implementation of the dust grain equilibrium
temperature and emission SED calculation in CUDA for the Monte-Carlo
radiation transfer code \mcrx \ \citep{jonsson06sunrise,
jonssonetal09}. The CUDA version obtains a speedup of more than two
orders of magnitude compared to a modern CPU core when using an Nvidia
Tesla C1060 high-performance computing card. We start by outlining the
calculation and the CUDA implementation in Sections~\ref{section_calc}
-- \ref{section_impl}, present accuracy tests and benchmarks in
Section~\ref{section_results}, and conclude with a discussion and
summary in Sections~\ref{section_discussion} \& \ref{section_summary}.

\section{The Calculation of Dust Emission}
 \label{section_calc}

\mcrx , a radiation-transfer code using the Monte-Carlo method to
calculate the transfer of light through a dusty medium, has been
described in previous papers. The basic computational algorithm was
described in \citet{jonsson06sunrise}, and the computation of dust
emission, along with results from an application of the code to
hydrodynamic simulations of spiral galaxies, in \citet{jonssonetal09}.
\mcrx \ is free software and the source code, including the CUDA
implementation used here, is available on the \mcrx \ home
page.\footnote{The \mcrx\ home page is \url{http://sunrise.familjenjonsson.org}.}
The calculation of the dust emission spectra, detailed in e.g.
\citet{misseltetal01}, is reviewed here to set the stage for
implementing this calculation in CUDA.

In the simplest calculation of emission from dust grains, each  grain
is simply assumed to emit like a modified blackbody, the temperature of
which can be found by equating the emitted luminosity with the
luminosity $L_h$ heating the grain (normally assumed to result from
absorption of shorter-wavelength radiation, but in principle from any
source, for example heating by high-energy electrons). This leads to
the equation 
\begin{equation}
\label{equation_equilibrium} L_h = \! \int \! \sigma_a ( \lambda ) B (
\lambda , T_e ) \,\mathrm {d} \lambda = 2 hc^2 \! \! \int \! \! \frac {
\sigma_a ( \lambda ) } { \left ( e^{ hc / ( k \lambda T_e ) } - 1
\right ) \lambda^5 } \,\mathrm {d} \lambda \> ,
\end{equation}
 which needs to be solved numerically to find the equilibrium
temperature $T_e$. In this formula, $\sigma_a ( \lambda )$ is the
(wavelength-dependent) absorption cross section of the dust grain, $B (
\lambda , T )$ is the blackbody function, and $c$, $h$, and $k$ are the
speed of light and the constants of Planck and Boltzmann.  Once the
equilibrium temperature is known, the dust emission spectrum $L_{
\lambda , e }$ can be calculated as 
\begin{equation}
L_{ \lambda , e } ( \lambda ) = \sigma_a ( \lambda ) B ( \lambda , T_e
) \>.
\end{equation}

Because a solution to Equation~\ref{equation_equilibrium} is necessary
for a number of dust grains of different sizes and compositions in a
number of different grid cells with different intensities of the
heating radiation, Equation~\ref{equation_equilibrium} needs to be
solved on the order of hundreds of millions of times to calculate the
dust emission in one of the simulation snapshots to which \mcrx \ is
typically applied. To show this more explicitly, the subscripts $s$ and
$c$ are used to indicate that the quantities depend on the dust grain
species (size and composition) and grid cell, respectively. The
calculations necessary can then be summarized by the following
equations: 
\begin{eqnarray}
\label{equation_part1} L_{ h ; c , s } & = & \int I_c ( \lambda )
\sigma_{ a ; s } ( \lambda ) \,\mathrm {d} \lambda \\
\label{equation_part2} L_{ h ; c , s } & = & 2 hc^2 \! \! \int \! \!
\frac { \sigma_{ a ; s } ( \lambda ) } { \left ( e^{ hc / ( k \lambda
T_{ e ; c , s } ) } - 1 \right ) \lambda^5 } \,\mathrm {d} \lambda \\
\label{equation_part3} L_{ \lambda , e ; c , s } ( \lambda ) & = &
\sigma_{ a ; s } ( \lambda ) B ( \lambda , T_{ e ; c , s } )
\end{eqnarray}

The need for a numerical solution to Equation~\ref{equation_part2},
using on the order of 1000 wavelength bins, and the fact that a global
iterative procedure is necessary to find the equilibrium distribution
of radiative intensities in cases where the dust is optically thick to
its own radiation \citep[see][]{jonssonetal09}, means that calculating
the dust emission involves the evaluation of at least $10^{ 11 } - 10^{
12 }$ exponentials. Evaluating an exponential is one of the most
computationally costly mathematical operations to perform on a modern
CPU, and this explains the large computational cost of the calculation.
Indeed, in a typical \mcrx \ run, the time for the dust emission
calculation outweighs the time for the transfer of radiation by a large
factor.  (For the case used to produce the results in
Section~\ref{section_results}, the temperature calculation takes $\sim
90 \%$ of the total time, and this generally increases further for
cases with more complicated geometry and higher optical depth.) Since
operations with high floating-point intensity are good candidates for
porting to a GPU due to their large advantage in raw floating-point
power, this indicates that performing the calculation on the GPU has a
large potential speedup.

The seasoned computationist will at this point argue that this
reasoning may be correct but ultimately irrelevant, as the sensible way
of calculating $T_e$ many times is by setting up a lookup table of $T_{
e ; s } ( L_{ h ; s } )$, thus bypassing the numerical solution of
Equation~\ref{equation_part2}. This is true (and is how the calculation
is actually performed in \mcrx ), but proceeding with the GPU
implementation is useful, for several reasons. First, as will be shown
below, the exact calculation of $L_e ( \lambda )$ performed on the GPU
\emph{is actually faster than a temperature table interpolation
implemented on the CPU}, a remarkable fact in itself. Second, the
implementation of the calculation of \emph{equilibrium} temperature
emission serves as a useful warm-up for implementing the calculation of
grains with \emph{fluctuating} temperatures \citep[as outlined in
e.g.][]{rajadraine89}, which is even more computationally intensive.

\section{The CUDA Programming Model}
 \label{section_cuda}

The CUDA programming model consists of functions, called
\emph{kernels}, that are meant to execute on the GPU. When the host CPU
starts a kernel, it is executed simultaneously for a large number of
threads. These threads are divided into \emph{blocks}, where threads in
the same block can communicate through a comparatively small amount of
shared, high-speed memory. On the current generation of hardware (CUDA
compute model 1.3), a block can consist of up to 1024 threads. Threads
in different blocks cannot communicate and are completely independent.

The actual hardware consists of an array of execution units called
Streaming Multiprocessors. Each multiprocessor executes one (or
several) thread blocks concurrently. The set of threads that executes
simultaneously, called a \emph{warp}, on current hardware is 32. Within
a warp, the threads execute one common instruction at a time, but with
individual data.  Out of the currently executing threads, the hardware
schedules warps with zero overhead, so warps that are unable to execute
because they are waiting for loads from the (uncached) global memory
will yield the hardware to other warps that are ready to execute.

The performance concerns for writing GPU programs are enumerated in the
CUDA ``Best Practices Guide'' \citep{cudabest2.3}.  Two of the most
important concerns of an efficient GPU algorithm are: first, minimizing
access to the (uncached) global memory by staging data in the fast
shared memory to avoid unnecessary loads and making sure loads are
\emph{coalesced}. For a coalesced load, all threads in the two halves
of a warp perform their loads in one memory transaction (unlike
uncoalesced loads which require several -- up to 16 -- memory
transactions and can thus take up to 16 times longer). The criteria for
coalesced loads vary between device generations, and the C1060 hardware
used here can coalesce all loads where threads access memory within the
same 128-byte segment (for loads of 4-byte data).  Because of the large
potential performance impact of uncoalesced loads, care has to be taken
when designing the memory access pattern of the kernels.

The second important concern is the hiding of the global memory latency
by maintaining high concurrency \citep{cudaguide2.3}. If the number of
concurrent warps executing is high enough, warps waiting for global
memory can always be substituted for others that are ready to execute.
The number of blocks that can execute concurrently on a multiprocessor
is determined by the number of processor registers and the amount of
shared memory used by the kernel, so minimizing use of these resources
is important for maintaining high concurrency.

\section{The CUDA Temperature Calculation}
 \label{section_impl}

With the above background, it is now clear that the dust temperature
calculation is well suited for a GPU implementation. The calculation
consists of millions of identical, independent evaluations of
Equations~\ref{equation_part1} -- \ref{equation_part3}. Each CUDA
thread will calculate the temperature and emission SED of a particular
dust grain species in a particular grid cell.

If the integrals in Equations~\ref{equation_part1} --
\ref{equation_part3} are replaced with the finite-difference equivalent
sums that will actually be calculated, the result is 
\begin{eqnarray}
\label{equation_sum1} L_{ h ; c , s } & = & \sum_l I_{ c , l } \sigma_{
a ; s , l } \, \Delta \lambda_l \\ \label{equation_sum2} L_{ h ; c , s
} & = & 2 hc^2 \! \sum_l \frac { \sigma_{ a ; s , l } \, \Delta
\lambda_l } { \left ( e^{ hc / ( k \lambda_l T_{ e ; c , s } ) } - 1
\right ) \lambda_l^5 } \\ \label{equation_sum3} L_{ \lambda , e ; c , s
, l } & = & \sigma_{ a ; s , l } B ( \lambda_l , T_{ e ; c , s } )
\end{eqnarray}
 where $l$ is the index used for the wavelength-dependent quantities.

In addition, after calculating the emission spectra of the individual
grains in a grid cell, the total emission in a grid cell needs to be
calculated by summing over the different grain species in the cell.
Combining this sum with Equation~\ref{equation_sum3} yields 
\begin{eqnarray}
\label{equation_sum4} L_{ \lambda , e ; c , l } & = & \sum_s L_{
\lambda , e ; c , s , l } n_s m_{ d ; c } \\ \label{equation_sum5} & =
& \sum_s \frac { 2 hc^2 \sigma_{ a ; s , l } n_s m_{ d ; c } } { \left
( e^{ hc / ( k \lambda_l T_{ e ; c , s } ) } - 1 \right ) \lambda_l^5 }
\>.
\end{eqnarray}
 where $n_s$ is the number of dust grains of species $s$ per unit mass
of dust, and $m_d$ is the mass of dust in the grid cell.

The calculation is performed with 4 kernels: The first kernel
precomputes the quantity $\sigma_{ a ; s , l } \, \Delta \lambda_l$,
which is used in Equations~\ref{equation_sum1} \& \ref{equation_sum2}.
The second kernel calculates the heating luminosity $L_{ h ; c , s }$.
The third solves Equation~\ref{equation_sum2} for the equilibrium
temperature $T_{ e ; c , s }$. The fourth kernel calculates $L_{
\lambda , e ; c , l }$ using Equation~\ref{equation_sum5}.

The thread blocks have specific sizes, such as 16 elements of $s$ and 8
elements of $l$, and the actual number of cells, dust species and
wavelengths may not be evenly divisable by the block size. For this
reason, the problem is padded with zeros so no edge checks are
necessary in the kernels.

All GPU calculations use single-precision floating point numbers. While
the GT200 architecture does have double-precision support, the
double-precision performance is a small fraction of that using
single-precision. As shown in Section~\ref{section_results}, the use of
single precision does not significantly affect the computed results.
However, care must be taken to avoid under- or overflowing the limited
dynamic range, as astrophysical numbers can span ranges rivaling that
of single-precision numbers.

The GPU also has a fast intrinsic function for calculating exponentials
in one clock cycle, {\tt \_\_expf}. This function has lower accuracy
than the math library function {\tt expf}, but is much faster. It was
verified that using {\tt \_\_expf} had no adverse impact on the
accuracy of the SED calculation, but resulted in about $40 \%$ higher
overall performance, so the results presented here use {\tt \_\_expf}
for all exponential calculations.

\subsection{Kernel 1}

The first kernel is trivial. Each thread is assigned an index of $( s ,
l )$, loads its elements of $\sigma_{ s , l }$ and $\Delta \lambda_l$,
and saves the product.  This calculation is not affected by the number
of cells and always takes negligible time.

\subsection{Kernel 2}

The second kernel calculates the heating luminosity $L_{ h ; c , s }$.
Looking at Equation~\ref{equation_sum1}, it can be seen that this
operation actually is a matrix multiplication of $( \sigma \Delta
\lambda )_{ s , l }$ and $I_{ c , l }$. Each thread is assigned an
index of (c,s) and calculates one element of $L_{ h ; c , s }$. To
minimize loads from global memory, the threads in the block first
stages blocks of $( \sigma \Delta \lambda )_{ s , l }$ and $I_{ c , l
}$ for an interval in $l$ in shared memory. Each thread then performs
the partial sum in Equation~\ref{equation_sum1} for the interval in $l$
that has been staged. The threads then continue to stage another
interval in $l$ until the full range in $l$ has been processed. The
matrix multiplication example is extensively discussed in the CUDA
Programming Guide \citep{cudaguide2.3}.

\subsection{Kernel 3}

The third kernel, which solves for the equilibrium temperature $T_{ e ;
c , s }$ is the most complex and time-consuming one. Each thread is
assigned an index $( c , s )$ and, since grains with the same $s$ in
different grid cells can share the same cross-section data, it is
advantageous for each block to process one grain species over many grid
cells. This is only possible as long as the entire cross section vector
for one grain species can fit in the 16kb of shared memory, which
currently limits the maximum number of wavelengths to $\sim 4000$.
(This limit is of no practical consequence at this time.)

Each thread then loads its value of the heating $L_{ h ; c , s }$ and
calculates a Newton-Raphson iterative solution of
Equation~\ref{equation_sum2}. The iteration is stopped when the energy
non-conservation in Equation~\ref{equation_sum2} is below a specified
fraction, currently 0.1\%.

The number of iterations required is a strong function of the accuracy
of the initial guess for the grain temperature, and because the
temperature calculation dominates the total computational time (both on
the GPU and CPU) this determines the total time of the calculation. To
determine a good guess, the relation between $L_h$ and $T_e$ was
studied, for both graphite, silicate and PAH grains. It was found that,
for these types of grains, a function of the form 
\begin{eqnarray}
\label{equation_guess} \log T & \approx & k + \beta_1 x + \beta_2 x^2
\mbox { , where } \\ x & = & \log L_h + \alpha_1 \log a + \alpha_2
\log^2 a \> ,
\end{eqnarray}
 where $a$ is the grain size, can be used to provide the initial guess.
Fitting to the grain cross sections resulted in parameters $k = 1.86$,
$\beta_1 = 0.189$, $\beta_2 = 3.41 \times 10^{ - 3 }$, $\alpha_1 = -
1.39$, and $\alpha_2 = 0.126$ (in SI units). With
Equation~\ref{equation_guess} as the initial guess, the number of
iterations required to obtain the temperatures of graphite and silicate
grains were at most 7, for any grain temperature below 1000K. The PAH
grains sometimes require more iterations, but no more than 9.  The
exception was that if Equation~\ref{equation_guess} suggested a
temperature of less than 5K, 5K was used. This is due to the
convergence behavior of the iteration, which can be unstable at low
temperatures if the initial guess is lower than the equilibrium
temperature.

Since the number of iterations required to obtain a solution of
required accuracy may vary between the threads in the block, the
threads may \emph{diverge}. In these cases, execution of the divergent
threads is automatically serialized by the hardware. In this case, the
threads that require fewer iterations will automatically pause and wait
for the thread that takes the longest. It is thus advantageous to make
sure the initial guess is good enough that the number of iterations
required for the different threads in a warp are all roughly similar,
but there is no need to handle this divergence explicitly in the code.

\subsection{Kernel 4}

The final part of the calculation is the generation of the total dust
emission SED in the grid cells from the equilibrium temperatures using
Equation~\ref{equation_sum5}.  Because this kernel uses many different
quantities, it has the most complicated shared-memory staging. Each
thread is indexed by $( c , l )$ and calculates $L_{ \lambda , e ; c ,
l }$. The sums over $s$ are then done by staging blocks of $\sigma_{ s
, l }$, $T_{ e ; c , s }$, and $n_s$ for the interval in $s$ covered by
the block in shared memory. The layout of this calculation is thus
largely similar to that of kernel 1. After the sum is completed, the
final quantity is obtained by multiplying by $m_{ d ; c }$. (Since each
value of $m_{ d ; c }$ is only used once, there is no point in staging
it in shared memory.)

After the kernels have finished executing, the resulting SED array is
copied back to system memory and, for the purpose of the test presented
here, compared with the SEDs calculated on the CPUs.

\section{Results}
 \label{section_results}

The results presented here compare the CUDA calculation, performed on
an Nvidia Tesla C1060 high-performace computing card with 4GB of memory
using CUDA version 2.3, to that performed on an 8-core Intel Xeon E5420
($2.5 \GHz$) Linux machine with 32 GB RAM. The code used for the
measurements in this paper is an improved version of \mcrx \ version
3.01\footnote{The code can be retrieved from the \mcrx\ SVN repository under {\tt svn/branches/cuda-paper}, revision 2400.},
and the CUDA calculation is fully integrated into \mcrx \ version 3.02.
To make the comparison more fair, the CPU code was tuned to improve
cache performance and load balancing, and to use the improved initial
guess in Equation~\ref{equation_guess}, though there are surely still
improvements that could be made.  The radiation intensities and dust
masses were taken from the simulation of the Sbc galaxy presented in
\citet{jonssonetal09}, with a grid containing 590k cells. The dust
grains were modeled using the graphite cross sections of
\citet{laordraine93} with 81 size bins and the size distribution of the
Milky-Way $R = 3.1$ model from \citet{draineli07}.  The wavelength grid
contained 968 wavelengths distributed from the Lyman limit to $1000
\um$.

\subsection{Accuracy}

\begin{figure} \begin {center} \includegraphics*[width=\columnwidth]{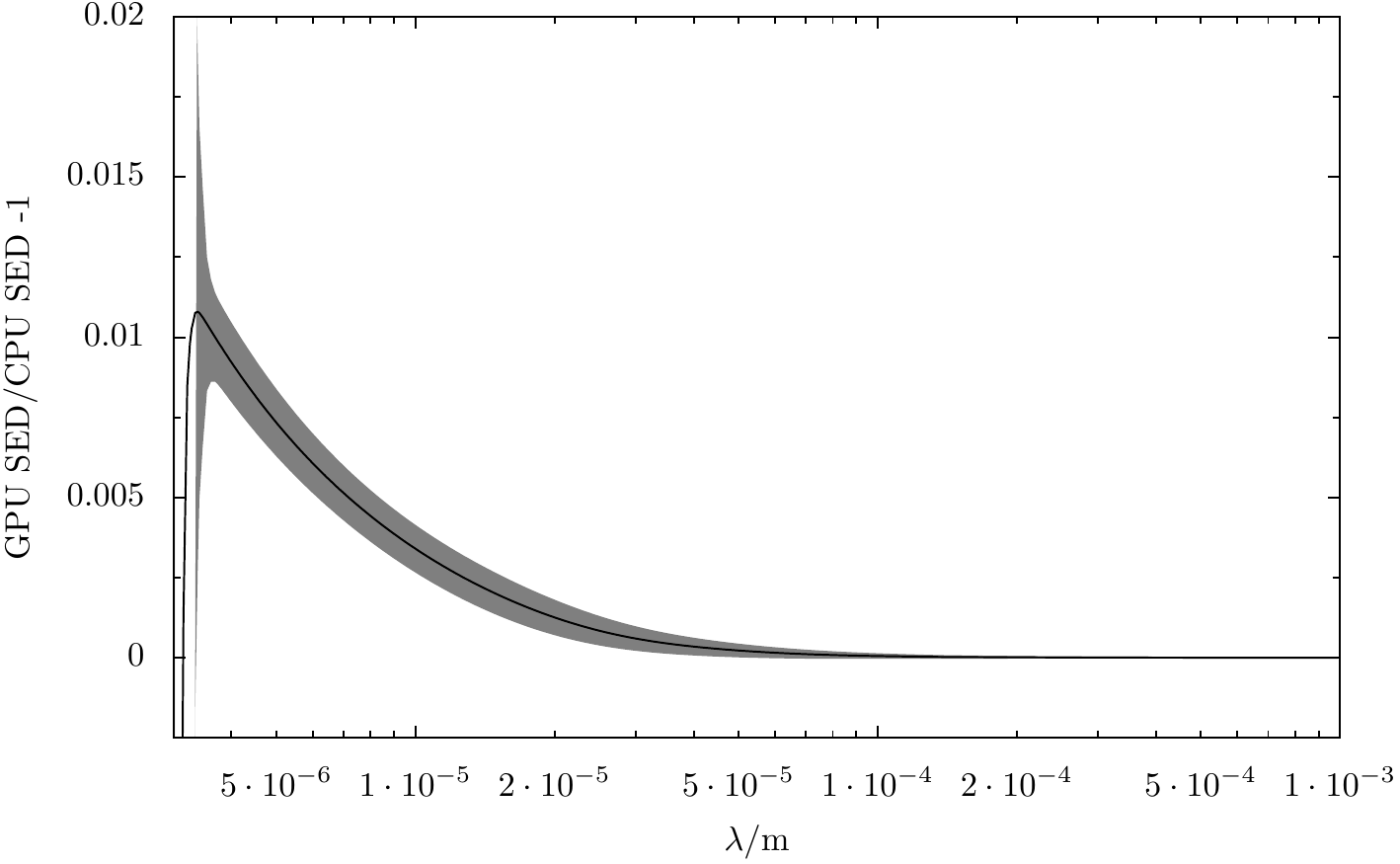} \end {center} \caption{ \label{plot_sedsigma} The dust emission SED calculated by the GPU compared to that calculated on the CPU. The line indicates the mean over all cells of the SED ratio, weighted by the CPU SED. The shaded region indicates the $1\sigma$ variance (also SED-weighted) over the cells. The blow-up at short wavelengths occurs because the GPU SED is calculated in single-precision and, for the dust temperatures encountered in the problem, the exponential blackbody cutoff will underflow to zero at short wavelengths. This increase in variance at short wavelengths has no practical significance as it occurs precisely because the actual emission drops to negligible levels.  } \end{figure}

To verify that the results of the CUDA calculation are correct, they
were compared against the extensively tested calculation in the
standard version of \mcrx \ performed on the CPU. The result is shown
in Figure~\ref{plot_sedsigma}, which shows the mean and variance, over
all grid cells, of the relative SED calculated on the GPU and CPU.
Because the SED changes dramatically from one grid cell to another, the
comparison was done by weighting the cells by the SED (as calculated on
the CPU) in that cell. If this is not done, the large variation in the
exponential cutoff at short wavelengths will cause the variance to be
dominated by the cells with cold dust -- precisely the cells that do
not contribute significantly to the total emission. For wavelengths
longer than $20 \um$, where grains in thermal equilibrum can be
expected to have an important impact, the difference between the two
calculations is about 0.1\%, shrinking rapidly at longer wavelengths.
The difference increases at shorter wavelengths, reaching 1\% at $4
\um$ and then blowing up as the single-precision exponential cutoff
underflows. The $1 \sigma$ spread around the mean in the results is
always negligibly small.

\subsection{Performance}

\begin{table*} \begin{tabular}{rrrrrrrrr} No. of cells & GPU & Kernel 1 & Kernel 2 & Kernel 3 & Kernel 4 & CPU & Speedup & CPU (interpolation) \\ 32 & 0.72 & $5.7\cdot 10^{-5}$ & $1.4\cdot 10^{-4}$ & $6.4\cdot 10^{-3}$ & $2.7\cdot 10^{-4}$ & 0.83 & 1.2 \\ 64 & 0.74 & $5.7\cdot 10^{-5}$ & $1.5\cdot 10^{-4}$ & $6.7\cdot 10^{-3}$ & $4.4\cdot 10^{-4}$ & 0.84 & 1.1 \\ 128 & 0.73 & $5.8\cdot 10^{-5}$ & $2.4\cdot 10^{-4}$ & $7.6\cdot 10^{-3}$ & $8.1\cdot 10^{-4}$ & 0.90 & 1.2 \\ 256 & 0.71 & $5.8\cdot 10^{-5}$ & $3.8\cdot 10^{-4}$ & $1.2\cdot 10^{-2}$ & $1.5\cdot 10^{-3}$ & 1.0 & 1.4 \\ 512 & 0.78 & $6.0\cdot 10^{-5}$ & $7.0\cdot 10^{-4}$ & $2.0\cdot 10^{-2}$ & $3.0\cdot 10^{-3}$ & 1.9 & 2.4 \\ 1024 & 0.76 & $6.7\cdot 10^{-5}$ & $1.2\cdot 10^{-3}$ & $3.5\cdot 10^{-2}$ & $6.0\cdot 10^{-3}$ & 3.6 & 4.8 \\ 2048 & 0.80 & $8.1\cdot 10^{-5}$ & $2.4\cdot 10^{-3}$ & $6.8\cdot 10^{-2}$ & $1.2\cdot 10^{-2}$ & 7.1 & 8.9 \\ 4096 & 0.89 & $8.8\cdot 10^{-5}$ & $4.8\cdot 10^{-3}$ & 0.13 & $2.4\cdot 10^{-2}$ & 14 & 16 \\ 8192 & 1.1 & $8.7\cdot 10^{-5}$ & $9.4\cdot 10^{-3}$ & 0.26 & $4.7\cdot 10^{-2}$ & 28 & 26 \\ 16384 & 1.5 & $8.7\cdot 10^{-5}$ & $1.9\cdot 10^{-2}$ & 0.51 & $9.5\cdot 10^{-2}$ & 57 & 38 \\ 32768 & 2.2 & $8.7\cdot 10^{-5}$ & $3.7\cdot 10^{-2}$ & 1.0 & 0.19 & 116 & 52 \\ 65536 & 3.9 & $8.8\cdot 10^{-5}$ & $7.5\cdot 10^{-2}$ & 2.0 & 0.42 & 238 & 62 \\ 131072 & 7.5 & $9.1\cdot 10^{-5}$ & 0.15 & 4.0 & 1.4 & 478 & 64 \\ 262144 & 14 & $8.6\cdot 10^{-5}$ & 0.30 & 8.0 & 2.8 & 963 & 68 \\ 524288 & 28 & $8.8\cdot 10^{-5}$ & 0.60 & 16 & 5.5 & 1925 & 69 & 385 \\ \end{tabular} \caption{\label{table_times} Execution wall clock times, in seconds, of the GPU and CPU (on 8 processors) calculations for different problem sizes. The times are averages of 3 runs. (The kernel execution times were generally very stable, while the times that include data transfer and CPU showed larger variation, presumably due to OS background tasks. The large start-up cost of the GPU calculation is evident. These numbers were obtained for calculations of graphite grains, but the results were approximately the same if silicate and PAH cross sections were used.} \end{table*}

\begin{figure} \begin {center} \includegraphics*[width=\columnwidth]{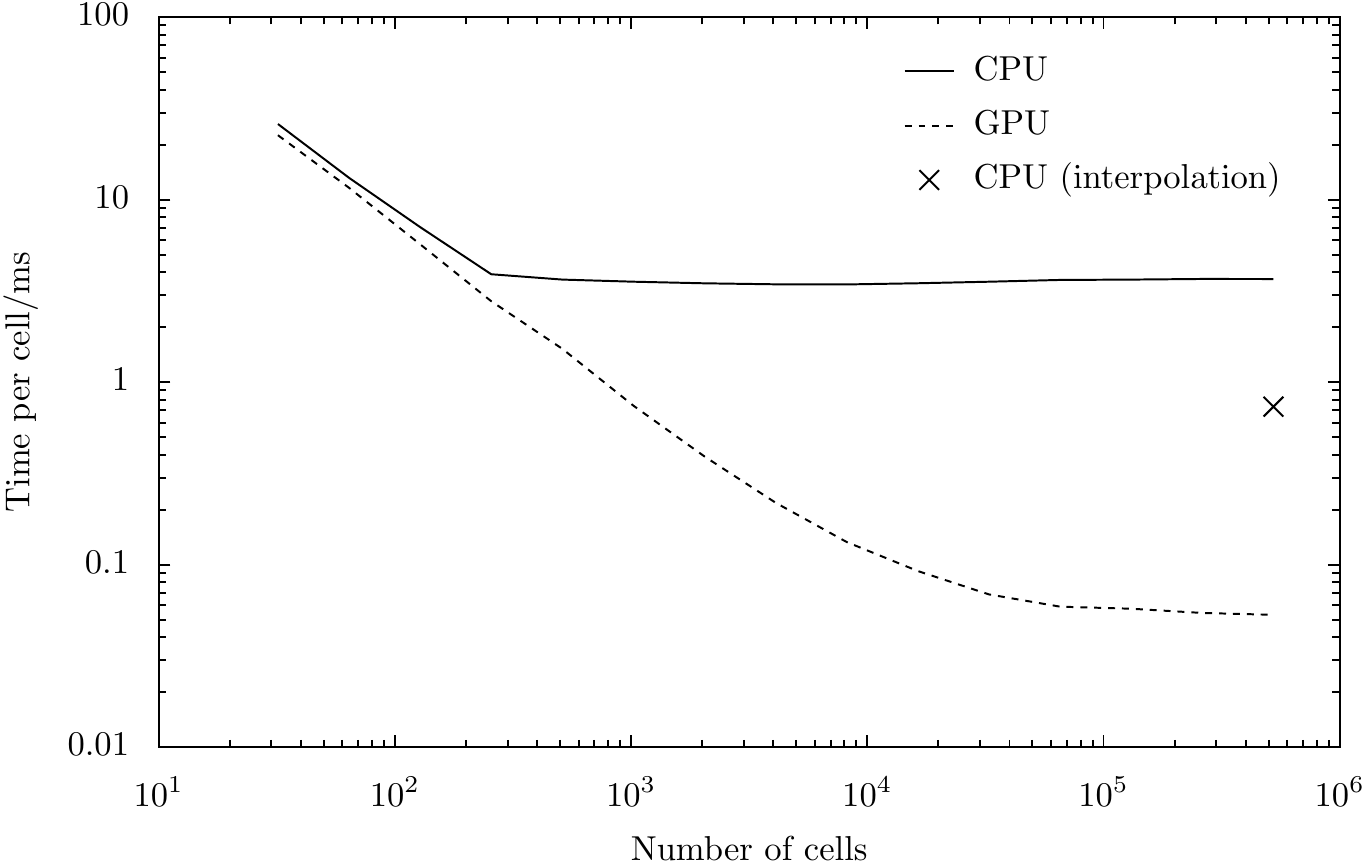} \end {center} \caption{ \label{plot_speedpercell} The wall clock time required to calculate the dust emission SED of one grid cell, as a function of the total number of cells, for the CPU (using 8 cores) and GPU. The time when interpolation was used on the CPU is indicated by a point, for the largest number of cells only.  This problem used the graphite cross sections of \citet{laordraine93}, with 81 size bins and 968 wavelengths.  For small problem sizes, the GPU time is dominated by kernel launch and data transfer overhead and is essentially independent of number of cells up to $>10^3$ cells. For the CPU calculation, a block size of 32 cells was used, so for 256 cells or less, there are not enough blocks to load all 8 cores, so the time is then independent of problem size. This is merely an artifact of the block size used.} \end{figure}

To collect performance measurements, the timers in the CUDA runtime
library were used to measure the execution times of the different
kernels, the total CUDA time (including all kernel runs and data
transfer), and the CPU calculation. These data are shown in
Table~\ref{table_times} and Figure~\ref{plot_speedpercell}. For large
problem sizes, the GPU calculation is 69 times faster than the CPU
calculation (on 8 cores). The GPU execution time is dominated by Kernel
3, the numerical temperature solution, which takes 57\% of the total
time. The time used by the 4 kernels adds up to only 79\% of the total
GPU time, the remaining time being data transfer to and from the
device.

Using the CUDA Visual Profiler (v2.3), the occupancy and memory
bandwidth of the kernels were determined. Kernels 1 \& 2 have full
occupancy, while kernels 3 and 4 have occupancies of 0.5. The
time-dominating kernel 3 is not very sensitive to the occupancy, as it
is not memory-bandwidth limited. The global memory bandwidth used by
kernel 3 is only about $95 \, { \rm { MB } / \rm { s } }$, far below
the maximum of $102 \, { \rm { GB / s } }$. (This is not surprising,
given the fact that the only per-cell quantities used by kernel 3 are
$L_h$ and $T_e$.)

The instruction throughput of kernel 3 is about 0.7 instructions per
cycle, indicating that the threads are not waiting for memory access
for a significant amount of time. The throughput would presumably be
even higher in the absence of divergent warps. According to the
profiler, about $4 \times 10^{ - 5 }$ of all branches are divergent for
this kernel.

\section{Discussion}
 \label{section_discussion}

It is clear from the performance numbers presented here that the GPU
temperature calculation dramatically outperforms the calculation done
on an 8-core fairly modern CPU. This is not very surprising, given that
the theoretical maximum floating point performance of the Tesla unit
used here is about 6 times greater than that of the 8 Xeon cores. What
is more surprising is that the difference in performance actually is an
order of magnitude greater than this.

While the large performance difference shows that the calculation
performed here is perfectly suited for the massively parallel GPU, one
part of the explanation is surely that the CUDA code is a low-level
calculation purposely written directly to conform to the rules for
getting maximum performance out of the GPU, while the CPU C++ code is
written at a much higher abstraction level that largely trusts that the
compiler can generate efficient code from the Blitz++ expression
template matrix library \citep{blitz} and that the CPU cache machinery
can hide memory latency. The performance comparisons presented here
thus say less about the innate performance difference between the two
architectures and more about the speedups that are possible when a
small part of an existing code is rewritten for maximum performance. It
is likely that the performance of the CPU code could be further
improved by rewriting the CPU calculation to the same low level as the
CUDA code, meticulously paying attention to cache performance and
blocking the loops to make sure cache thrashing is avoided. The virtue
of the GPU memory access scheme, however, is that the rules are simple
and straightforward. Knowing exactly how caches, prefetchers, etc.,
work on any given CPU is much more difficult, but even if cache
performance on the CPU was optimal, there is no way the CPU calculation
could outperform the GPU. This is especially true given the heavy use
of exponentials in the calculation, since the GPU can calculate an
exponential in a single clock cycle.

As mentioned earlier, the \mcrx \ temperature calculation in production
runs is actually done using a linear interpolation table, containing
2000 temperature points between $3 \,$K and $1500 \,$K. The time
required is shown in the final column of Table~\ref{table_times}.
Remarkably, calculating the SED on the GPU is \emph{still more than an
order of magnitude faster} than when interpolating the temperature on
the 8 CPU cores. This is not quite as unexpected as it might seen,
however. Even if the temperature is obtained in zero time, it is
necessary to first calculate the heating rate. Subsequently, the
emission SED must be calculated using Equation~\ref{equation_sum5}.  It
can be seen from Table~\ref{table_times} that, on the GPU, the
temperature calculation (kernel 3) takes about 70 percent of the total
execution time. For the CPU code, this fraction is 80 precent, so
completely eliminating the temperature calculation can only speed up
the CPU calculation by maximally a factor of 5. The actual reduction in
time by using the interpolation on the CPU is indeed very close to 80
percent, which leaves the GPU still more than an order of magnitude
faster.

One interesting point is that once a problem has been formulated to fit
into the CUDA programming model, scaling to future generations of
hardware is virtually ensured. As the calculation already has been
subdivided into independent blocks, these blocks can simply be
distributed across a larger number of multiprocessors with essentially
perfect scaling.

How does the experience here apply to the (much more computationally
intensive) calculation of thermally fluctuating grains? In contrast to
the calculation here, calculating the temperature probability
distribution of thermally fluctuating grains essentially requires
inverting a matrix for each grain species \citep{rajadraine89}, the
size of which is determined by the number of temperature levels in the
distribution. If each thread still calculates the temperature
distribution of a specific cell and grain species, shared memory use
will increase from storing $\sigma_{ a ; s , l }$ to also storing the
full transition matrix between the temperature levels for that grain
species.  Storing the elements of the temperature probability
distribution during the calculation will also require more thread-local
storage, potentially increasing the use of bandwidth to global memory.
More work needs to be done to determine the best way to implement this
calculation on the GPU, and this will be presented in a future paper.

\section{Summary}
 \label{section_summary}

We have presented an implementation of the calculation of dust grain
equilibrium temperatures in CUDA, and compared the performance of this
implementation running on a GPU with that performed on a normal
multicore CPU. The GPU vastly outperforms the 8 CPU cores, with a
factor of 69 speedup. This is almost two orders of magnitude faster
despite the fact that the difference in theoretical maximum
floating-point performance is only a factor of 6, showing that the
inherently parallel, exponential-heavy nature of the calculation is
perfectly suited to the GPU. As grain temperature calculations, not the
actual transfer of radiation, are normally the most computationally
expensive part of calculating the SED of a galaxy, this holds great
promise for accelerating such calculations.

\section{Acknowledgements}

The Nvidia Professor Partnership Program kindly donated a Tesla C1060
in support of this project, and we thank David Luebke, Chris Henze, and
Chris Hayward for discussions about how to port various parts of \mcrx
\ to GPUs. We also thank the anonymous referee for constructive
comments that helped both the clarity of the presentation and the
efficiency of the calculation. PJ was supported by Spitzer Theory Grant
30183 from the Jet Propulsion Laboratory and by programs HST-AR-10958
\& -11758, provided by NASA through grants from the Space Telescope
Science Institute, which is operated by the Association of Universities
for Research in Astronomy, Incorporated, under NASA contract
NAS5-26555.

\bibliographystyle{elsarticle-harv} 
\bibliography{patriks} \label{lastpage} \end {document}

%% file: natlatex_macros.tex
\renewcommand{\Ref}{\ref}

\newcommand{\um}{\, {\mu \rm m}}

\newcommand{\GHz}{\, {\rm GHz}}

\newcommand{\mcrx}{\textit{Sunrise}}